\documentclass{aa}  
\usepackage{color}
\usepackage{xcolor}
\usepackage{ulem}
\usepackage{graphicx}
\usepackage{txfonts}
\usepackage{amsmath}
\usepackage{caption}

\usepackage[hypertexnames=true]{hyperref}
\begin{document}

   \title{The torsion of stellar streams}

   \subtitle{ and the overall shape of galactic gravity's source}

   \author{Adriana Bariego--Quintana
          \inst{1}
          \and
          Felipe J. Llanes--Estrada\inst{2}
          }

   \institute{IFIC-Univ. Valencia, c/ Catedr\'atico Jos\'e Beltr\'an, 2, E-46980 Paterna, Valencia, Spain \\
              \email{adriana.bariego@gmail.com}
         \and
             Universidad Complutense de Madrid, IPARCOS \& Dept. Física Teórica, Plaza de las Ciencias 1,
             28040 Madrid, Spain\\
             \email{fllanes@ucm.es}
             }

 
  \abstract
   {Flat rotation curves $v(r)$ are naturally explained by elongated (prolate) Dark Matter (DM) distributions, and we have provided competitive fits to the SPARC database. To further probe the geometry of the halo, or the equivalent source of gravity in other formulations, one needs observables \textcolor{black}{outside the galactic plane}. Stellar streams, poetically analogous to airplane contrails, but caused by tidal dispersion of massive substructures such as satellite dwarf galaxies, would lie on \textcolor{black}{their own} plane (consistently with angular momentum conservation) should the DM-halo gravitational field be spherically symmetric. \textcolor{black}{Tracks resembling entire orbits} are seldom available because their periods are commensurable with Hubble time, with streams often presenting themselves as short segments. }
   {Therefore, we aim at establishing stellar stream torsion, a local observable that measures the deviation from planarity in differential curve geometry, as a diagnostic providing sensitivity to aspherical DM distributions which ensures the use of even relatively short streams.
 }
   {We perform small-scale simulations of tidally distorted star clusters to check that indeed a central 
force center produces negligible torsion while distorted haloes can generate it. 
Turning to observational data, we identify among the known streams those 
that are at largest distance from the galactic center and likely not affected by the Magellanic clouds, as most 
promising for the study, and by means of polynomial fits we extract their differential torsion.  }
   {We find that the torsion of the few known streams that should be sensitive to most of the Milky Way's DM
Halo is much larger than expected for a central spherical bulb alone. This is consistent with nonsphericity of the halo.}
   {Future studies of stellar stream torsion with larger samples and further out of the galactic plane should
    be able to extract the ellipticity of the halo to see whether it is just a slight distortion of a spherical shape
    or rather resembles a more elongated cigar.}


   \maketitle
%
\section{Introduction: shape of Dark Matter Haloes}\label{sec:intro}
  
The problem of galactic rotation is the empirical statement that rotational velocity around the galactic center
seems to flatten out for a large fraction of the galaxy population where this has been measured at long enough distances~\citet{1978ApJ...225L.107R}. 
\textcolor{black}{This creates a puzzle if the data is interpreted in terms of rotation external to a spherical source around which orbital equilibrium (Kepler's third law written for the velocity $v$) demands falling $v$ for stellar objects or clouds of gas, }
\begin{equation} \label{Keplers3}
\frac{v^2}{r} = \frac{GM}{r^2} \implies v=\sqrt{\frac{GM}{r}}\ .
\end{equation}
\textcolor{black}{The most accepted explanation is the existence of additional Dark Matter or a modification of the law of motion under gravity.}

Because typical velocities in spiral galaxies are of order 200-300 km/s, $v/c\sim 10^{-3}$, 
relativity is a correction and Newtonian mechanics should get the bulk of the rotation right. 
Therefore, either a modification of mechanics, such as MOND~\citet{1983ApJ...270..365M}, or a modification of the gravity 
source, typically in the form of a spherical Dark Matter halo~\citet{1985Natur.317..595F}, are invoked. 
MOND however runs into problems at larger, cosmological scales~\citet{2001ApJ...561..550A,2006PhRvL..97w1301D}; and a spherical DM distribution 
has to be fine-tuned to have very nearly an isothermal $\rho(r)\propto 1/r^2$ profile to explain 
the flatness of the rotation curve.

If we inhabited a two-dimensional cosmos, however, the natural gravitational law would be $|F|\propto \frac{1}{r}$
instead of $\propto \frac{1}{r^2}$ and the observed rotational law would be $v_{\rm 2D}\propto {\rm constant}$ which is the law that the experimental data demands.
We do not; but a cylindrical matter source achieves the same dimensional reduction by providing translational symmetry
along the $OZ$ symmetry axis of the cylinder~\citet{2010arXiv1009.1113S,2021Univ....7..346L}. If the linear density of the cylindrical dark matter source is 
$\lambda$, we can write
\begin{equation}
\frac{v^2}{r} = \frac{2G\lambda}{r}\implies v = \sqrt{2G\lambda}\ .
\end{equation}

That is, the constant velocity function $v(r)$ is natural for a filamentary source. Moreover, if 
the rotation curve is only measured to a finite $r$, obviously the case, the source does not need to be infinitely
cylindrical: it suffices that it be prolate (elongated) instead of spherical, as shown by detailed fits~\citet{2022arXiv220406384B}
to the SPARC database~\citet{2016AJ....152..157L}
and consistently with simulations of DM haloes~\citet{2006MNRAS.367.1781A}.

Observables in the galactic plane alone, such as detailed rotation curves, cannot distinguish between competing 
models such as spherical haloes with nearly $\rho_{\rm DM}(r)\propto \frac{1}{r^2}$ profiles or elongated haloes with arbitrary profile. To lift the degeneracy between shape and profile one needs to find adequate, simple observables from  data \textcolor{black}{outside the galactic plane}.

For a while now, stellar streams~\citet{2007SciAm.296d..40I} in the Milky Way galaxy have been a promising new source of information on the DM distribution~\citet{2023ApJ...954..195N}, as they will eventually be for other galaxies~\citet{2022ApJ...941...19P}, \color{black}
or more generically, for the effective gravitational potential~\citet{2014MNRAS.443..423S}. 
Streams have been proposed~\citet{10.1093/mnras/stw1957} to detect  DM clumps (for example, presumed subhaloes) which may collide with a stream and leave a gap imprinted therein

\color{black}
In the rest of this article we develop what we think is a key observable to be measured on those streams to bear on the question of the overall shape of the presumed halo. 
Section~\ref{sec:notorsion} is dedicated to reviewing the definition of torsion in differential curve geometry, \textcolor{black}{because it is the parameter which locally quantifies the aplanarity of a curve in three dimensions} and to showing
that, around a spherical halo, orbits as well as streams are torsionless.
\textcolor{black}{(Note the distinction: streams can be thought of as a bundle of nearby orbits, one for each constituent body. If a whole stream is taken as a perfect orbit, biases in the full reconstruction of the gravitational potential can creep in~\citet{2013MNRAS.433.1813S}.)
}
Section~\ref{sec:elongationtorsion} then shows how
we expect tidal streams around elongated gravitational sources  to show torsion if there is a component of the velocity parallel to the axis of elongation of the source. Section~\ref{sec:streams} makes a reasonable selection among the known stellar streams in our galaxy and we plot the torsion calculated along each of them, showing that there seems to be a signal here. Section~\ref{sec:conclusion} then discusses how further studies can improve the conclusion.

\section{Orbits and streams around central potentials are torsionless}\label{sec:notorsion}
\subsection{Torsion quantifies separation from orbital planarity}

Before explaining why we wish to propose torsion as a useful observable to probe the DM halo, to fix notation,  let us recall a few concepts of differential geometry.
There~\citet{2017Carmo} the torsion of a curve measures how sharply it is twisting out of the osculating plane,
instantaneously defined by the velocity and normal acceleration. 

To a curve ${\bf r}(t)$ in three-dimensional space parametrized by an arbitrary variable $t$ 
we can associate an arclength $s(t)=\int^t \arrowvert {\bf r}(t') \arrowvert dt'$ 
and the tangent vector ${\bf T} = d{\bf r}/ds $;  if at a certain point $P$ the curvatuve is non-zero,  
then the normal vector at $P$ is defined by ${\bf N} = d{\bf T}/ds $ (its inverse modulus giving the radius of the circumference best approximating the curve at $P$); and the binormal vector (that completes the Frenet-Serret trihedron)  
by the vector product of both,
\begin{equation}
	{\bf B} = {\bf T}\times {\bf N}\ .
	\label{eq:binormal}
\end{equation}

If the curve is perfectly planar the tangent and normal vectors will always lie in the same plane,
and in such case the binormal vector stays parallel to itself along the curve. Any natural definition of torsion
will then yield zero.

But if the curve twists out of  \textcolor{black}{its osculating} plane (like a uniformly advancing helix, which corresponds to constant torsion), 
the binormal vector will acquire a rotation.
Torsion will then measure the speed of that rotation of the binormal, and it is a locally defined vector at 
each point $P$ along the curve ${\bf r}(t)$, as the scalar product of the intrinsic derivative of ${\bf B}$
and the normal vector (this discounts the change of the modulus of ${\bf B}$ and rather measures its twisting),
\begin{equation}
	\tau = -  \frac{d{\bf B}}{ds }\cdot {\bf N} \ .
	\label{eq:torsion}
\end{equation}
If the arc length is not at hand and the arbitrary parameter $t$ needs to be used, then a convenient formula (with the prime denoting $d/dt$) is
\begin{equation}\label{torsionfromders}
\tau = \frac{({\bf r}'\times {\bf r}'')\cdot {\bf r}'''}{|{\bf r}'\times {\bf r}''|^2} \ .
\end{equation}
Since up to three derivatives of the position along the curve need to be computed, several adjacent points of a discretized curve are needed to extract the torsion: but it is still quite a local observable that does not need long trajectory stretches. 

We are going to demonstrate the use of this observable $\tau$ for stellar streams, particularly around the Milky Way, 
to determine the shape of the gravitational potential of its DM Halo.

\subsection{Movement around a Newtonian spherical source}\label{subsec:centralVtraj}

Newtonian gravity predicts, for motion around a spherical body, 
\begin{equation} \label{Newtonsphere}
{\bf r}'' = \frac{\bf F}{m} = -GM \frac{\bf r}{|{\bf r}|^3}
\end{equation} 
with $M$ the mass inside the sphere of radius $|{\bf r}|$. 
The needed third derivative can be computed in a straight-forward manner, taking into account that 
$|{\bf r}|' = \hat{\bf r}\cdot {\bf r}'$ is the modulus of the projection of the velocity 
along the visual from the origin,
\begin{equation}\label{3rdderivative}
{\bf r}''' = \frac{-GM}{|{\bf r}|^2}\left( 
\hat{\bf r}'-\frac{2}{|{\bf r}|}(\hat{\bf r}\cdot {\bf r}')\hat{\bf r}
\right) 
\end{equation}
in terms of components along the velocity and along the position.

Because of Eq.~(\ref{Newtonsphere}), 
\begin{equation}
({\bf r}'\times {\bf r}'') \propto ({\bf r}'\times {\bf r})
\end{equation}
and therefore, observing that both terms of Eq.~(\ref{3rdderivative}) are proportional to either
${\bf r}'$ or ${\bf r}$, we see that $({\bf r}'\times {\bf r}'')\perp {\bf r}''' $. 
Therefore, the scalar product in the numerator of Eq.~(\ref{torsionfromders}) vanishes, 
and thus $\tau=0$ for motion around a spherical body.

The planarity of the orbit around a central potential is, of course, a textbook consequence~\citet{2001Goldstein} of the 
conservation of the direction of the angular momentum vector $\hat{\ \bf L}$ that in this language
is parallel to the binormal vector. And additionally, the Newtonian gravity law is not strictly necessary: any central potential will yield the same result. This observation is of particular interest for the MOND explanation of the galactic rotation curves in Eq.~(\ref{sec:intro}) since, while the intensity of the acceleration induced by matter is different from Newtonian mechanics, the central direction of the force is respected: MOND likewise predicts no torsion.

\begin{figure*}[h!]
  \centering
    \includegraphics[width=.88\linewidth]{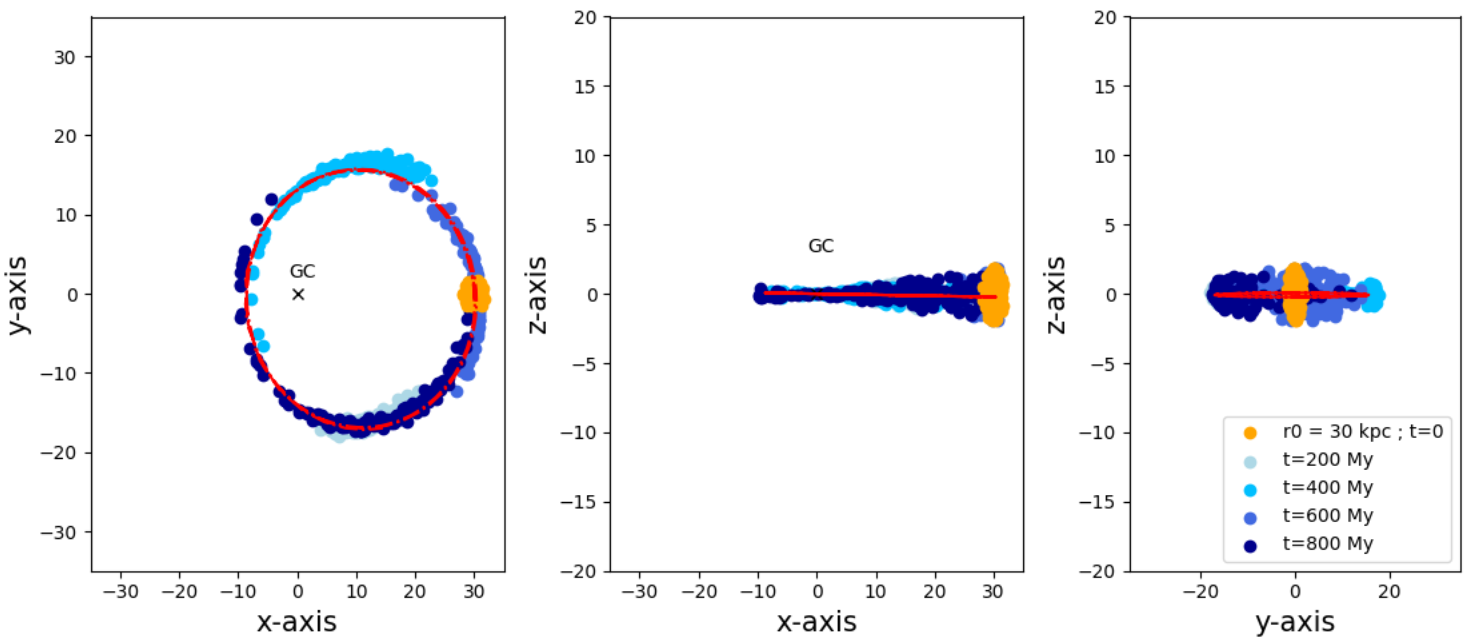}
  
  \includegraphics[width=.4\linewidth]{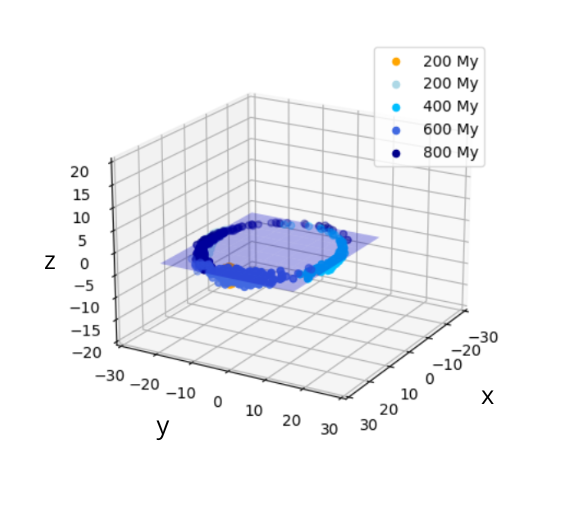}
  \includegraphics[width=.48\linewidth]{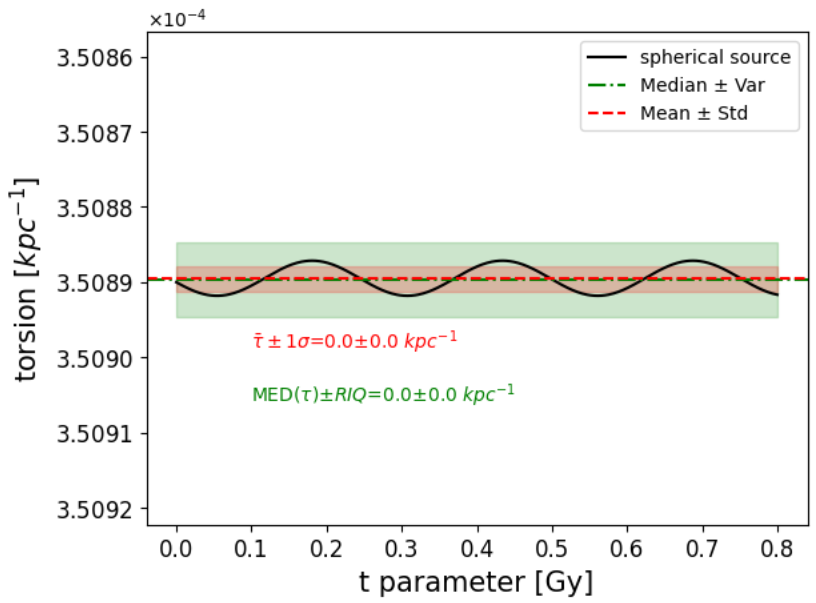}

  \caption{ 
\color{black} Top row: simulated movement of a 200 star cluster around a spherical gravitational source with parameters akin to those of the Milky Way, in the three planar projections, extracted at times $t=0$ $My$, $t=200$ $My$, $t=400$ $My$, $t=600$ $My$ and $t=800$ $My$. 
Lower left: three-dimensional view with the containing plane. The initial cluster has a radius of 2 kpc and is at a distance of 30 kpc from the galactic origin of coordinates. Member stars have slightly different velocities $v_* = 220 {\rm kpc} \pm \Delta v$ due to the additional random $\Delta v$. 
(All axes in kpc.)
Lower right: extracted torsion as function of time, from an analytical helicoidal fit to the resulting stream with $\chi^2/N_{gdl} = 0.86$, which yields a value of $\tau\simeq 3.5\times 10^{-4}$ kpc$^{-1}$ (a one-star orbit would return exactly zero torsion).
}
  \label{fig:cloudspherical}
\end{figure*}

\subsection{Simulation of an N-point stellar stream around a spherical gravitational source}
\label{subsec:simulation1}

The discussion presented in Subsec.~\ref{subsec:centralVtraj} refers to the torsion of one test body under a central field. If the body lost dust grains forming a kind of contrail, its shape through space would be a planar curve. 
But stellar streams are not quite of this nature, rather the result of the tidal stretching of a globular cluster or dwarf galaxy~\citet{2019RMxAA..55..273N}. Since each  object in the cluster starts off at a different height $z$ with respect to the galactic plane, its orbit around the center of force lies on a slightly different plane, the effect being 
that the cluster, in addition to stretching, contorts, with the upper particles passing under the center of mass (cm) to become the lower ones with each half orbit. 

We here show that this effect is negligible and the torsion of a stream around a central potential can safely be neglected, as the cm of the stream follows one of the trajectories of subsection~\ref{subsec:centralVtraj}, with $\tau=0$.
Rather than entangling the discussion in detailed theory, a couple of simple simulations will  illustrate the point.

We simulate a globular cluster of $N$ \textcolor{black}{(specifically 100, 200, 500, 700, and 1000 particles)} pointlike stars, randomly distributed at $t=0$ over a sphere of radius $R_0$ (of order one or few kpc, typically; in the following simulation, 2 kpc) at a distance $|r|$ from the galactic center (of order 30 kpc  in the following example),  with a certain mass $m^*_i$ and common initial velocity ${\bf v}_*\sim 220$ kpc. An additional random velocity kick $\Delta v_0 = Gm_{\rm cluster}/2r_0$ in a random direction is given. 
We then let it evolve under the gravitational force of the central source with mass $M\in(10^{9}, 10^{12})M_\odot $, standing 
for a galactic bulb or a spherical DM halo, and we allow for a correction due to the inner binding forces of the cluster.
This is small because the random masses are taken in the interval $m_*\in(0, 20) M_\odot$ and thus their mutual interactions are orders of magnitude smaller than those with the galactic center.
The constant $GM$ of the central source can conveniently be eliminated in terms of the typical velocity of circular orbits around the galactic center, from orbital equilibrium $v_{\rm rot}^2/r = GM/r^2$. For the Milky Way this is typically 220 km/s.

We show the simulation in Fig.~\ref{fig:cloudspherical}. The concentrated orange points 
\textcolor{black}{in the stream visualization plots}  (three dimensional views as well as Cartesian projections as marked in the axes) mark the initial cluster at $t=0$. The evolved cluster at later times, the clouds of cyan, blue, and purple dots, is seen to stretch under tidal tensions. 
\textcolor{black}{
The stream remains near the (slightly tilted) plane containing the initial velocity, without developing motion outside that plane, thus with negligible torsion for each individual orbit.
Nevertheless, a measurable $\tau\sim O(10^{-4})$ is visible on the lower-right plot, due to the need to fit a curve to the dispersed, criss-crossing points.
}

\clearpage
\section{Orbits around elongated potentials and $\tau \neq 0$} \label{sec:elongationtorsion}

\subsection{Movement around a Newtonian cylindrical source}\label{subsec:cylVtraj}
We now move on to  quickly show  how torsion is expected to look for an orbit around a perfectly cylindrical source of gravity, 
in a discussion paralleling that of subsection~\ref{subsec:centralVtraj}. We naturally employ cylindrical coordinates
$(\rho,\varphi,z)$, so that, \textcolor{black}{denoting the time derivative by a $'$ sign as in Eq.~(\ref{eq:torsion})}, 
\begin{eqnarray}
{\bf r}   &=& \rho \hat{\boldsymbol{\rho}} + z \hat{\bf z} \\
{\bf r}'  &=& \rho' \hat{\boldsymbol{\rho}} +\rho \varphi' \hat{\boldsymbol{\varphi}} + z' \hat{\bf z} 
\label{velcylindrical}\\
{\bf r}'' &=& ( \rho'' -\rho \varphi^{'2})  \hat{\boldsymbol{\rho}} +(2\rho'\varphi'+\rho \varphi'')\hat{\boldsymbol{\varphi}}
 + z'' \hat{\bf z} \label{acccylindrical}
\end{eqnarray}
where in the acceleration we recognize, from left to right, the radial, centrifugal, Coriolis, azimuthal and vertical accelerations, respectively. 
The force law is the same as that for a line of charge in electromagnetism, except of course with the constant replaced,
so that in terms of the linear mass density $\lambda$,
\begin{equation} \label{cylindricalforce}
{\bf r}'' = \frac{\bf F}{m}  = \frac{-2G\lambda}{\rho} \hat{\boldsymbol{\rho}} +0\cdot \hat{\boldsymbol{\varphi}}
+0\cdot \hat{\bf z}\ .
\end{equation}
Comparing with the general form in Eq.~(\ref{acccylindrical}) we recover $z''=0\implies z'={\rm constant}$ (reflecting  translational invariance along the  $OZ$ axis) and $\rho^2\varphi'={\rm constant}$ so that the third component $L_z/m$ of angular momentum per unit mass is conserved just as in the central force problem.  However, now the direction of ${\bf L}$ is not conserved, so that the binormal vector changes and one expects a torsion. To obtain it, start from Eq.~(\ref{cylindricalforce}) and take a further derivative to obtain
\begin{equation}\label{thirdder}
{\bf r}''' = \frac{-2G\lambda}{\rho} \left( -\frac{\rho'}{\rho} \hat{\boldsymbol{\rho}} +\varphi' \hat{\boldsymbol{\varphi}}
\right)
\end{equation}
(valid for the Newtonian force with cylindrical symmetry only).
Calculating the cross-product of Eq.~(\ref{velcylindrical}) and Eq.~(\ref{cylindricalforce}), while using the righthandedness of the trihedron $(\hat{\boldsymbol{\rho}},\hat{\boldsymbol{\varphi}},\hat{\bf z})$ to evaluate each basis vector product, yields
\begin{equation}
{\bf r}'\times{\bf r}'' = (-2G\lambda)\left[ -\varphi' \hat{\bf z} +\frac{z'}{\rho} \hat{\boldsymbol{\varphi}} \right]\ .
\end{equation}
Next we take the scalar product with ${\bf r}'''$ and evaluate Eq.~(\ref{torsionfromders}) to obtain the torsion, yielding
\begin{equation}\label{tcylinder}
\tau = \frac{z'\varphi'}{(\rho\varphi')^2+z^{'2}} = \frac{1}{\rho} \frac{v_zv_\varphi}{v_\varphi^2+v_z^2}
\end{equation}
that  we have cast in an easier to remember form in the second expression. Clearly, for there to be a torsion we need both azimuthal and vertical velocities (so stellar streams in the galactic plane are not sensitive, as expected). Additionally, because $|v_zv_\varphi| \leq v_z^2+v_\varphi^2$, torsion belongs to the interval $\tau\in [-\rho^{-1},\rho^{-1}]$,  
so its maximum magnitude is controlled by the distance from the stellar stream segment to the galactic axis.

\begin{figure}[h!]
  \centering
\includegraphics[width=.9\linewidth]{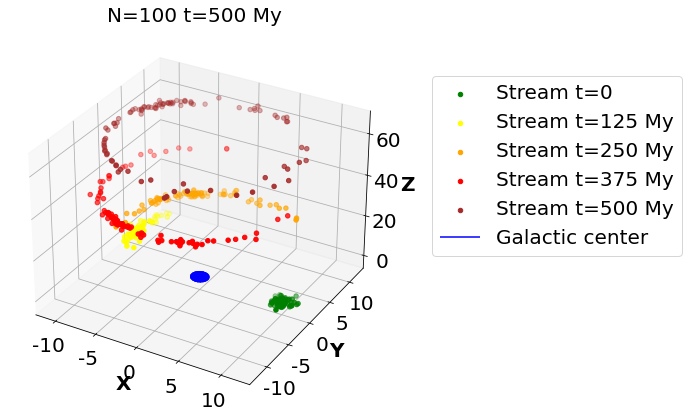}
  \caption{ We show the spread of a 100-body stellar cluster with initial vertical speed $v_z= 50 $km/s after $t=$ 0, 125, 250, 375 and 500 My.  The resulting stream at different times clearly shows an upward advancing helix moving around the z-axis. } 
  \label{fig:cyltor}
\end{figure}

\begin{figure*}
  \centering
  \includegraphics[width=.8\linewidth]{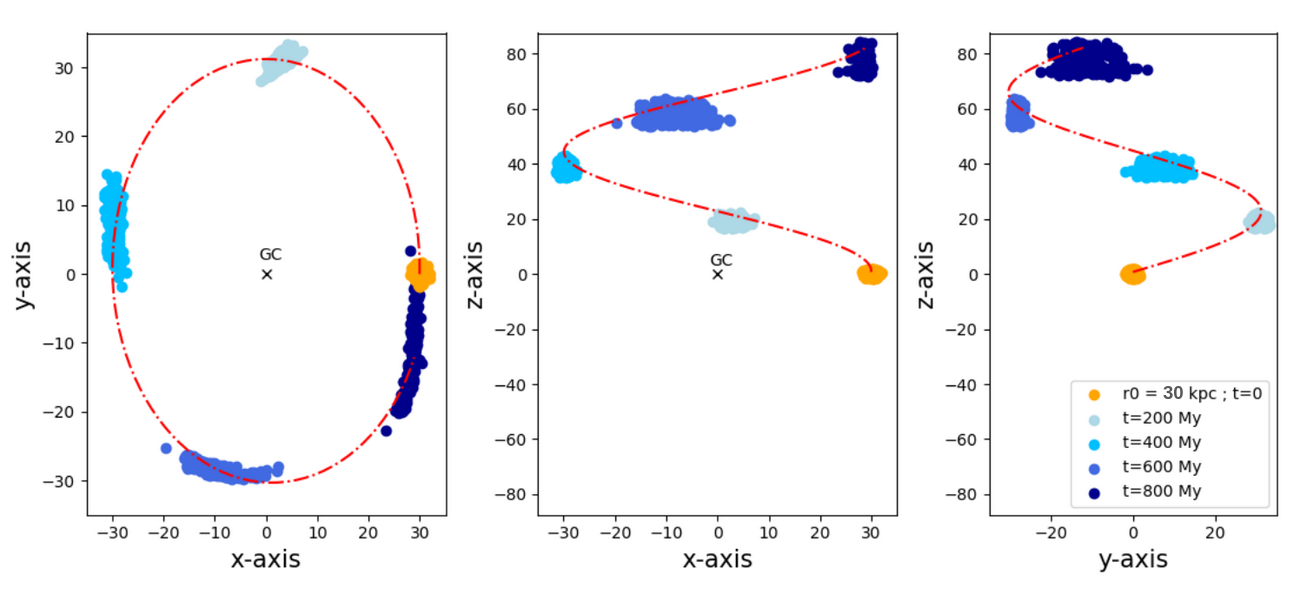}
  \includegraphics[width=.4\linewidth]{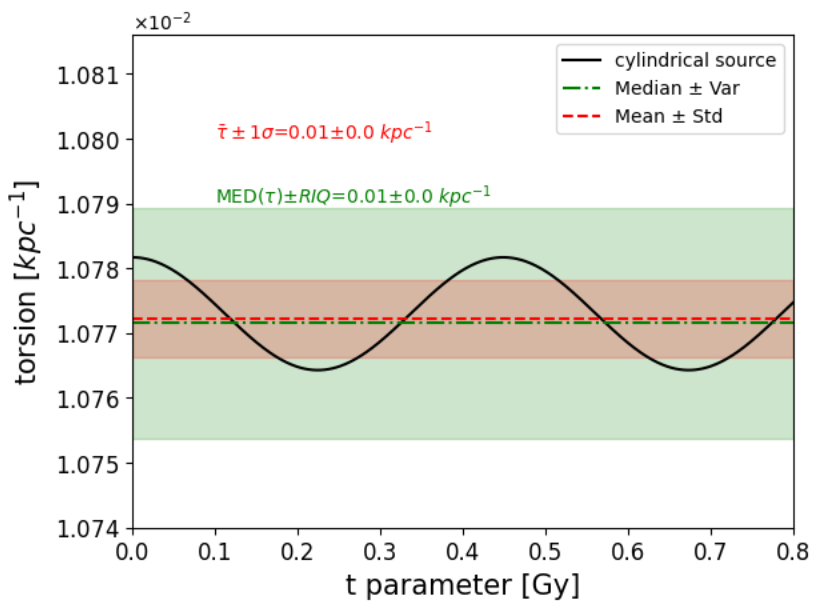}
  \caption{ \color{black}
The top row tracks the motion of a star ensemble around around a toy-model, perfectly cylindrical source with an initial $v_z=$ 90 km/s, with the same conventions as in Fig.~\ref{fig:cloudspherical}. A stream forms along a spiral trajectory. From a helicoidal fit to the cloud of points, as needs to be done on real data, with $\chi^2/N_{gdl}=1.16$ we find a value for the torsion of $\tau\simeq 1.1\times 10^{-2}$  kpc$^{-1}$. This is in agreement with the analytical estimate in the main text below Eq.(\ref{torsionestimate}). 
}
  \label{fig:cylindrical}
\end{figure*}

\subsection{Simulation of an N-point stellar stream around a cylindrical source}

Next we proceed to repeat the exercise of subsection~\ref{subsec:simulation1} with the same starting data, but replacing the central spherical Newtonian source by a cylindrical source. 
\textcolor{black}{The force, relegated to  Eq.~\ref{eq:cylacc2} in the appendix, now distinguishes the vertical  acceleration from its two components parallel to the galactic plane.}

Its first term is the acceleration caused by the cylindrical gravitational source 
(that along the $OZ$ axis being zero by translational symmetry).
Its linear mass-density $\lambda = M/L$ is obtained from the typical rotation curve around a galaxy $v_{rot} = \sqrt{2G\lambda}$~\citet{2021Univ....7..346L}. 
The second term of Eq.~\ref{eq:cylacc} is, again, the correction due to the tiny binding of the stellar streams stars among themselves, together with Eq.~(\ref{eq:cylacc2}). 
\textcolor{black}{
We show two typical computations in figures~\ref{fig:cyltor}, 
a three-dimensional view, and
~\ref{fig:cylindrical}, a more detailed analysis of a run.}

If the starting velocity profile was perfectly set in the $XY$ plane perpendicular to the cylinder, the torsion would still be zero as per Eq.~(\ref{tcylinder}). We give it a slight tilt and then the orbit starts behaving 
as a helix.
\color{black}This can be appreciated  by studying the originally compact cluster of stars 
 at different times up to  1 Gyr, 
seeing it spread and ascend in a spiral, showing a  torsion which is a factor 30 larger than for the stream around a purely spherical source. 

The value of the torsion extracted from the numerical simulation is $\tau\simeq 1.08 \cdot 10^{-2}$ kpc$^{-1}$.
This is in excellent agreement with the analytical estimate that we now present. To obtain it, we first note that Eq.~(\ref{tcylinder}) for the torsion, with the parametrization of the stream track given in Eq.~(\ref{ellipticalhelix}) and an almost circular planar projection (semiaxes satisfying $a\simeq b$) is
\begin{equation}
\label{torsionestimate}
\tau|_{\rm helix} \simeq \frac{c\omega}{a^2\omega^2+c^2}\ .
\end{equation}
Here, $c$ is the vertical ``velocity'' $v_z$ in terms of the arbitrary curve parameter $t$, and $\omega=v_\varphi/\rho$ the rotational ``angular velocity'' 
with $\rho=a$ the radius of the orbit in the $XY$ plane.
With the setup data for the simulation in  figure~\ref{fig:cylindrical}, namely c $\sim 80$ kpc/Gyr, $a\sim 30$ kpc (note that the curvature scale is then $a^{-1}\sim 0.03$ kpc$^{-1}$) and $\omega \sim 7$ rad/Gyr, 
Eq.~(\ref{torsionestimate}) yields $\tau|_{\rm helix} \simeq 0.011$ kpc$^{-1}$ in agreement with the numerical value 0.0108 extracted from the simulation.
The slight oscillation seen in figure~\ref{fig:cylindrical} is due to the ellipticity of the helix (with the oscillation period of about half a Gyr corresponding to half of the actual period of the trajectory because of the reflection symmetry of an ellipse).

Unfortunately for observing this in a physical setup, at a fixed observation time the stream appears quite planar, and it is only in a computer where we can simulate several time intervals at the same time that we see the helix.
This is because of the translation symmetry along the cylinder, which does not allow for a vertical tidal stretching of the stream.
On the upside, this makes the extracted torsion very insensitive to the number of bodies in the simulation, from $10^2$ to $10^3$, since the additional bodies maintain reasonably close orbits in the $z$--$v_z$ vertical phase space.
Providing an initially larger speed dispersion along the OZ axis increases the length of the stream but also blurs it, making its reconstruction more difficult.
Thus, we move onto the next case, in which both a sphere and a cylinder (a very idealized toy model of a galaxy and a prolate DM halo) respectively cause the vertical tidal stretching and the helicoidal motion.
\color{black}

\subsection{Sphere and cylinder with initial nonvanishing $ v_{z,0}$}
\label{section:zcomponent}

To close this section, we will combine together both types of sources, a sphere (akin to a visible-matter galactic bulb) 
and a cylinder (mimicking an elongated DM halo). 
For the sphere we take the typical mass of a galaxy $M_{s}\in (10^9, 10^{12})M_\odot$ \citet{2011ApJ...743...40B},  \citet{1987gady.book.....B}  and for the cylinder we use the expression for the linear mass density that we obtain from the asymptotic velocity at large $r$ in the rotation curve $v_{rot}$ of the Milky Way, as exposed in the previous section~\ref{sec:elongationtorsion}.

The updated expression for the acceleration of the stars in the stream is obtained by combining Eqs. (\ref{eq:accelerationspherical}) and (\ref{eq:cylacc}), that is,
\begin{eqnarray}
z'' &=& -\frac{v_0^2}{|{\bf r}|^3} z\\ 
{\bf r}_\perp'' &=& -\frac{|{\bf v}|^2}{|{\bf r}_\perp|^2 }{\bf r}_\perp -\frac{v_0^2}{|{\bf r}|^3 }{\bf r}_\perp
\label{cylsph}
\end{eqnarray}
where $v$ is taken from the galactic rotation velocity when it has flattened out at large $r$, and $v_0$ estimated from the visible mass.

We have added a small but appreciable (5km/s) contribution to the 
\textcolor{black}{average initial velocity in the $z$ direction, so that \\
\textbf{$v_*^i$} = $(220+\Delta v^i_x, \Delta v^i_y, 5 + \Delta v^i_z) $ km/s to induce sufficient vertical  motion that will
generate torsion as per Eq.~(\ref{tcylinder}). Additionally, there is a dispersion $\Delta v^i$ which depends on the particular star and is randomized from a distribution between 1 and 1000 times the quantity $\sqrt{G\sum_j m_{*j}/r_0}$
in terms of the mass of the small cluster and the characteristic length scale. There is no particular physical meaning to this initial dispersion except to be able to handle an ensemble of stars which are not all perfectly collinear to test the fitting procedure.
} 

\begin{figure*}[h]
  \centering
\includegraphics[width=.2\linewidth]{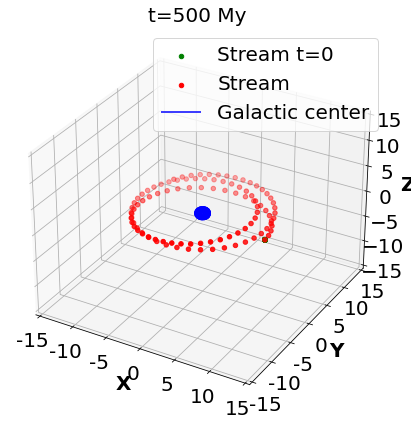}
\includegraphics[width=.2\linewidth]{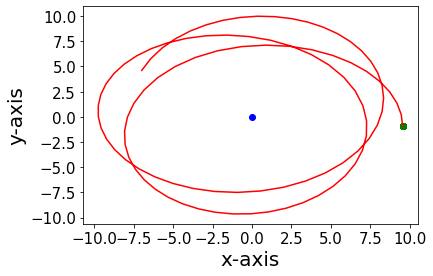}
\includegraphics[width=.2\linewidth]{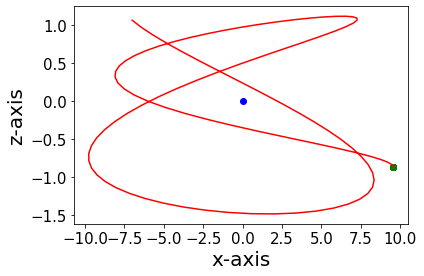}
\includegraphics[width=.2\linewidth]{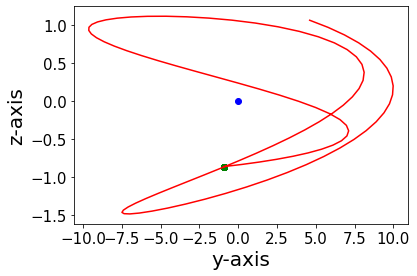}
\includegraphics[width=.2\linewidth]{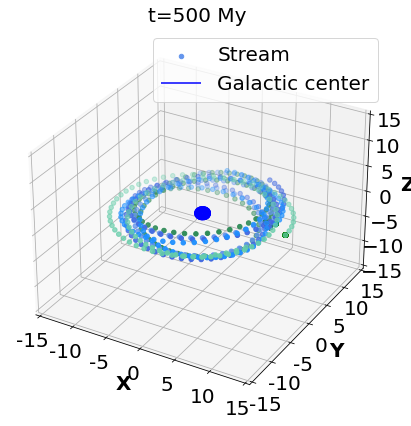}
\includegraphics[width=.2\linewidth]{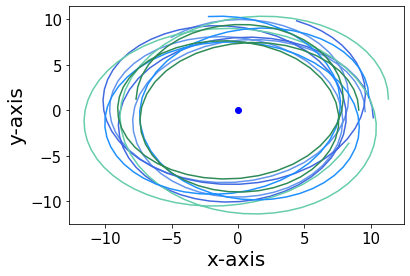}
\includegraphics[width=.2\linewidth]{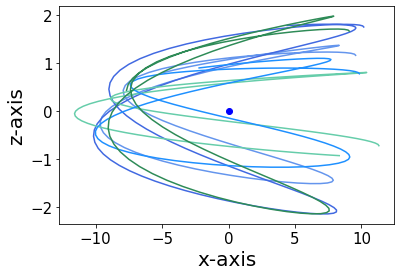}
\includegraphics[width=.2\linewidth]{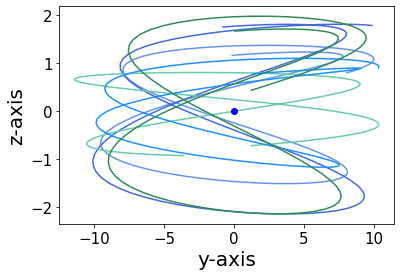}	
  \caption{The first row represents the trajectory of a point mass with, alternatively, initial vertical velocities $v_z=$1 and 2 km/s around a source combining both a sphere and a cylinder. The second row extends this to simultaneously show the motion of five stars.}
\label{fig:sphere+cloudcylinder}
\end{figure*}

\begin{figure*}[h]
  \centering
 \includegraphics[width=.8\linewidth] {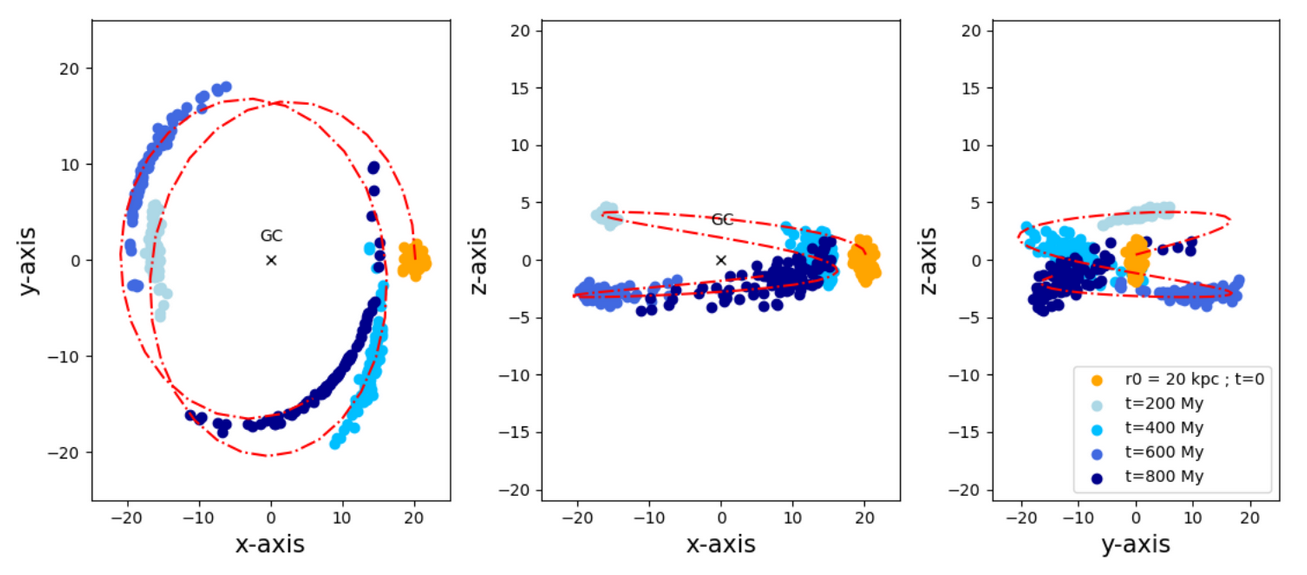}
 \includegraphics[width=.4\linewidth] {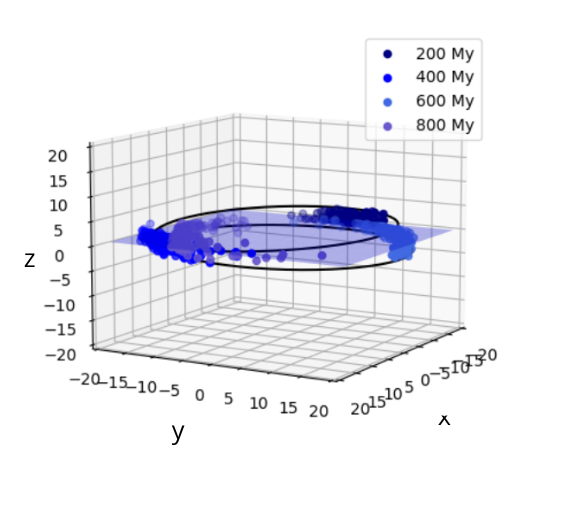}
 \includegraphics[width=.4\linewidth] {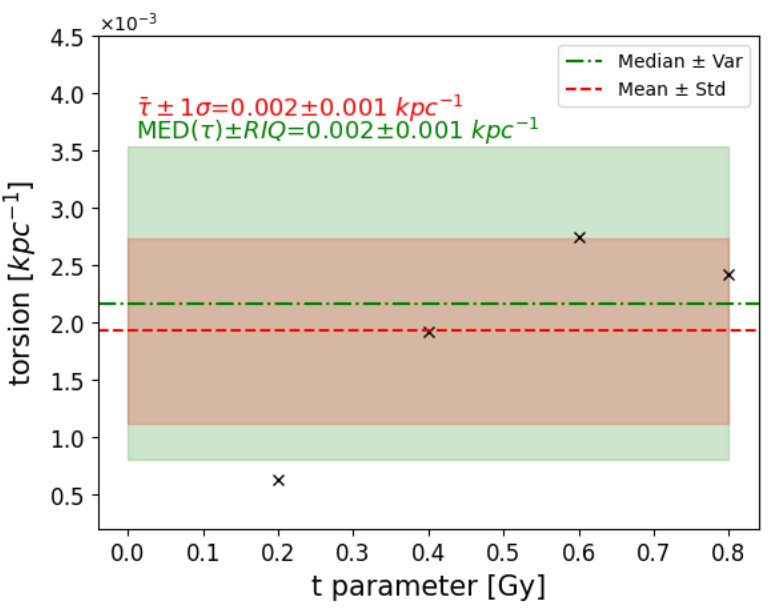}
   \caption{ Top row: Formation of a stellar stream from an initial cluster orbiting a combined spherical+cylindrical source \textcolor{black}{(a toy model representing the galaxy plus an elongated DM halo) 
   as in Fig.~\ref{fig:sphere+cloudcylinder}, with positions given at the times indicated in the legend.} \textcolor{black}{
   Lower left: three-dimensional view of the stream compared with a plane containing the origin and the initial stream's position and velocity, showing that at subsequent times, the stream leaves that plane. Lower right: piecewise helicoidal fit. A nonvanishing  torsion can be extracted for each of the short segments at different times, yielding a mean value of $\tau = 0.22\cdot 10^{-3}$ kpc$^{-1}$. }
   } 
  \label{fig:sphere+cloudcylinder2}
\end{figure*}


In Fig.~\ref{fig:sphere+cloudcylinder} we clearly observe traits of the motion around cylinder+sphere sources as described in~\citet{2021Univ....7..346L}. 
Along the symmetry $OZ$ axis, a star will describe harmonic oscillations between the two hemispheres due to the Newtonian pull of the spherical part of the distribution acting towards the center (unless it is provided with escape velocity, in which case it will approach an asymptotic trajectory, a helix around the $OZ$ axis).
The orbit on the $XY$ plane is not closed due to the additional $1/r$ force from the elongated DM halo. The net effect in three dimensions can be seen as a precession of the orbital plane around the $OZ$ axis, with the trajectory creating complicated helicoidal patterns.

The $N$-body simulation in Fig.~\ref{fig:sphere+cloudcylinder2} reflects this and clearly shows the appearance of nonvanishing torsion in the stellar stream. \textcolor{black}{From a piecewise helicoidal fit with Eq.~(\ref{ellipticalhelix}) to the stream at different times $t=200$ My, $t=400$ My, $t=600$ My and $t=800$ My, we obtain different values for the torsion above the value $\tau>10^{-4}$ kpc$^{-1}$ that we found on streams around a purely spherical source (basically, our irreducible background). 
The actual value happens to be apparently smaller than that for the purely cylindrical geometry presented in the previous section, but this is a contingent effect due to the initial conditions of the stream, and bears no significance. 
(The  $\chi^2/N_{gdl}$ values for each of the visible stream segments, one at each successive time, are $ 0.35, 0.32, 0.23$ and $1.88$, respectively.)
The three-dimensional rendering in the left bottom plot clearly shows the torsion causing an out of osculating plane movement.
}

\subsection{Torsion in a galaxy with a spherical halo and a galactic plane}
\label{subsec:suelo}

\begin{figure}[h]
\begin{center}
\includegraphics[width=0.6\linewidth]{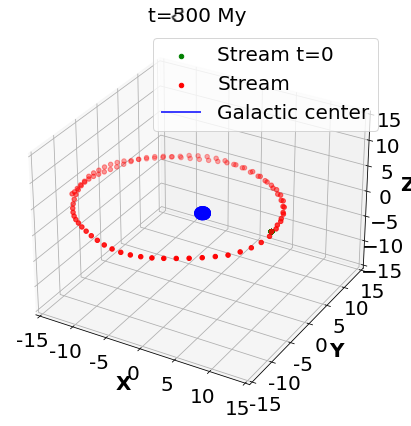}
\includegraphics[width=0.6\linewidth]{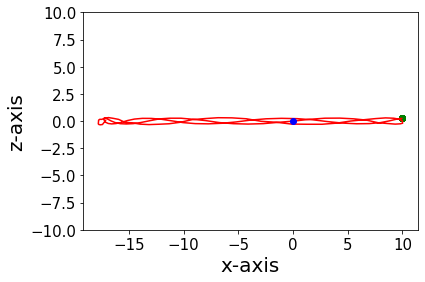}
\includegraphics[width=0.6\linewidth]{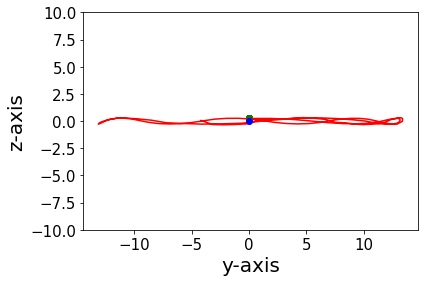} \end{center}
\caption{   Simulation of a small tidal stream in the presence of a typical non-spherical galactic source, here
the galactic plane model of Eq.~(\ref{sphereanddisk}). From top to bottom: three-dimensional view, $XZ$ and $YZ$ sections.
\label{fig:withdisk}
}
\end{figure}

We wish to have a reference for a minimum torsion that we would consider ``normal'' in order of magnitude, so that if extensive studies of stellar streams show that their torsion exceeds that level, one could reject the hypothesis of a spherical halo. 
For this purpose we propose here a toy model in which the halo is taken spherical, but we add a disk component. This adds a vertical (not radial) velocity outside of the galactic plane that points towards it. 

The simplest (and coarsest) such model takes the galactic disk as being uniform and infinite.
This is a reasonable approximation only for streams that do not elevate too much along the $OZ$ axis; otherwise it provides an upper bound to a more realistic torsion (since such additional vertical force will always be larger than that of a finite disk, whose effect will fall off with $z$).
In that case, observed torsions above this bound would still entail an incompatibility with a spherical halo to be studied further.

Therefore, in this minimum-torsion model we take the acceleration caused by the external force as
\begin{eqnarray} \label{accsphereplane}
a_{i=1,2}= &-&GM\frac{x^i}{(x^2+y^2+z^2)^{3/2}} \nonumber\\
a_3 =  &-&GM\frac{z}{(x^2+y^2+z^2)^{3/2}}  - 2\pi G \sigma    {\rm sign}(z)       
\ . \label{sphereanddisk}
\end{eqnarray}

In this equation, $\sigma$ is a surface mass density for the disk, in the range
$50-100 M_\odot/$parsec$^2$ which is a usual estimate~\citet{1989MNRAS.239..571K}, \citet{1991ApJ...367L...9K} at 8 kpc from the galactic center.

Fig.~\ref{fig:withdisk} shows the characteristic wobbling of movement near the galactic plane caused by the planar disk,
which is qualitatively consistent with \citet{2021MNRAS.503.2539C}.

We can provide an analytical estimate of the torsion following the now familiar reasoning. 
Since an instantaneous velocity that is parallel to any of the three coordinate vectors of the cylindrical base
$\{ \hat{\boldsymbol{\rho}}, \hat{\boldsymbol{\varphi}}, \hat{\mathbf{z}} \}$ will display zero torsion, we take a trajectory combining two of them,
\begin{equation}
{\bf r}' = v_\varphi \hat{\boldsymbol{\varphi}} + v_z \hat{\mathbf{z}}\ .
\end{equation}
Multiplying by the acceleration in Eq.~(\ref{accsphereplane}) we obtain
\begin{equation}
{\bf r}' \times {\bf r}'' = \frac{GM\rho}{r^3}(v_\varphi \hat{\mathbf{z}} - v_z \hat{\boldsymbol{\varphi}}) 
-Gv_\varphi \left(  \frac{M}{r^3}z+(2\pi)\sigma {\rm sign}(z) \right)\hat{\mathbf{\rho}}\ .
\end{equation}

To construct the determinant $(({\bf r}' \times {\bf r}'')\cdot {\bf r}''')$ necessary for the torsion, we evaluate the third derivative outside the galactic plane (over which it is undefined),
\begin{equation}
{\bf r}''' = -\frac{GM}{r^2}\left( \hat{\bf r}' - \frac{2}{r}(\hat{\bf r}\cdot \mathbf{r}')\hat{\mathbf{r}}
\right)
\end{equation}
that is in the plane given by position and velocity, employing
\begin{eqnarray}
\hat{\bf r}'  &=&  \frac{1}{r} \left( {\bf v} - (\hat{\bf r}\cdot {\bf v}) \hat{\mathbf{r}}
\right) \nonumber \\
&=& \frac{1}{r} \left( v_\varphi \hat{\boldsymbol{\varphi}} + v_z \hat{\mathbf{z}}
-\frac{zv_z}{r^2}(\rho  \hat{\boldsymbol{\rho}} + z \hat{\mathbf{z}}) \right) \ .
\end{eqnarray}
A slightly tedious but straightforward calculation then yields
\begin{equation}\label{tplanesphere}
\tau  =  \frac{-6\pi\, (M\sigma)\, (|z|\rho)\, (v_zv_\varphi)}
{v_\varphi^2(M^2 r +4\pi\sigma M|z|r^2+4\pi^2\sigma^2r^5)+v_z^2M^2\rho^2/r}\ .
\end{equation}

The numerator has mechanical dimensions of a squared momentum, and the denominator of squared momentum times length,
yielding the correct $1/L$ dimensionality of the torsion. Moreover, the structure of the numerator shows that in the
presence of a spherical source ($M$) alone, or a \textcolor{black}{galactic} plane ($\sigma$) alone, the torsion vanishes as it should. Likewise, both components of the velocity have to be nonvanishing as in Eq.~(\ref{tcylinder}) for the cylindrical source; and the torsion is null both on the galactic  plane ($z=0$) and on its perpendicular axis through the center of the sphere ($\rho=0$).
We can then numerically evaluate Eq.~(\ref{tplanesphere})  to obtain the floor value of the torsion that we should expect to be able to use in the galaxy. Taking into account that the galactic plane is not infinite so that the elevation $z$ will yield a diminishing multipolar field, it may be that galactic torsions from a spherical halo plus disk are even smaller; what we mean by this estimate is that those streams that may be found with larger values need to be further investigated
as they may be teaching us something about the dark matter halo or about dark matter inhomogeneities. 

Employing z$\sim 1$kpc, $\rho\sim r\sim 10$ kpc, $v_z\sim v_\varphi\sim 220$ km/s (to take the most conservative floor to the torsion), $M\sim 10^{12}\ M_\odot$, and $\sigma\sim 10^8\ M_\odot/$kpc$^2$ as already discussed, 
the denominator of Eq.~(\ref{tplanesphere}) is dominated by the $M^2$ terms, with the $\sigma r^2$ correcting $M$ only at the percent level. With these numbers we then find $\tau\sim -9\cdot 10^{-4}$kpc$^{-1}$.

We then conclude that torsions of stellar streams below $10^{-3}$ kpc$^{-1}$ in our galaxy can be explained without resort to deformed dark matter haloes or exotic phenomena. Of the few streams presently known at large radial distances, some present torsions  below this level and are thus
of no further interest for this application of the shape of the haloes. It is those that reach $\tau$ at the percent level that deserve further scrutiny to bear on the halo shape, among the ones known and in future searches for streams.

\section{Stellar streams in the Milky Way and their torsion} \label{sec:streams}
\begin{figure}[h!]
  \centering
  \includegraphics[width=1\linewidth]{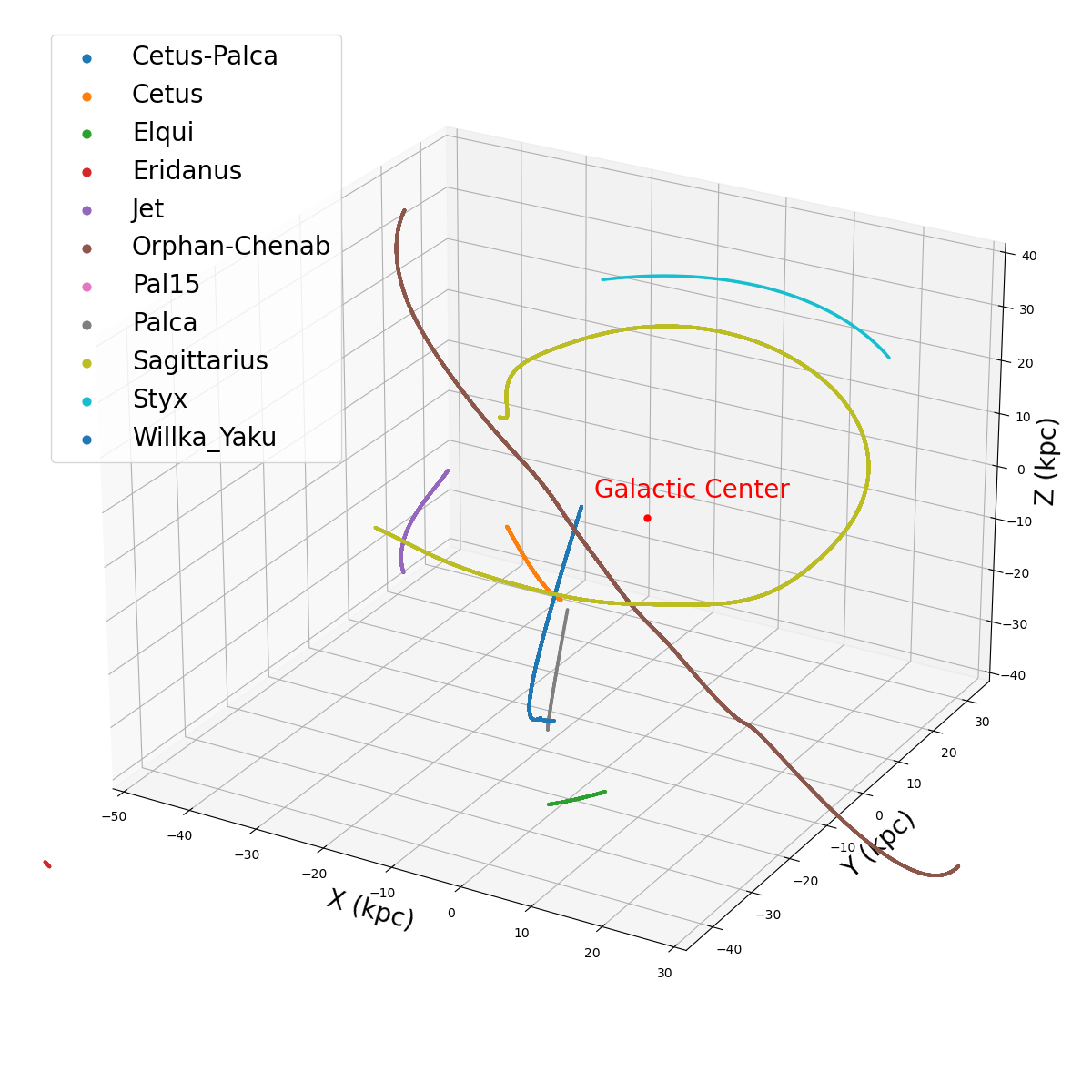}
  \caption{Trajectories in galactocentric coordinates for some of the MW outer streams, extracted from \citet{10.1093/mnras/stad321}. }
  \label{fig:MWstreamsGC}
\end{figure}

\begin{figure*}
  \centering
  \includegraphics[width=.9\linewidth]{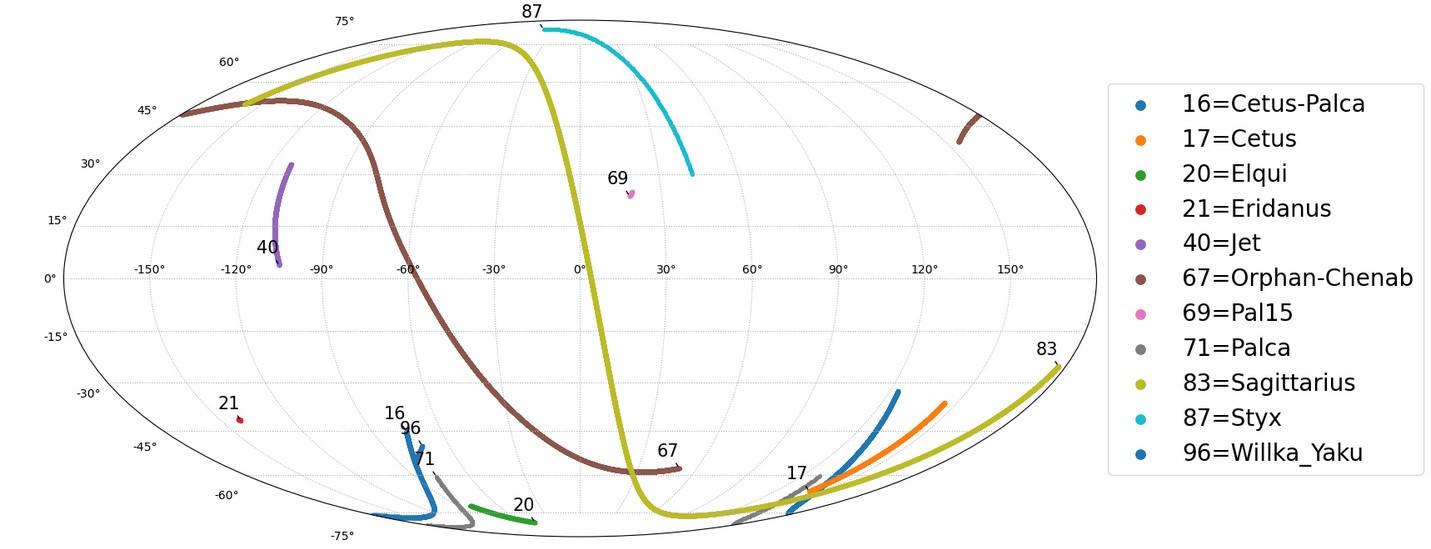}
  \caption{Trajectories for some of the MW outer streams  in Mollweide projection, extracted from \citet{10.1093/mnras/stad321}.}
  \label{fig:MWstreams}
\end{figure*}

In this section we finally turn to some of the known stellar streams in the Milky Way. 
We select as  relevant those found at distances $d>30$ $kPc$ from the galactic center, 
so that the internal structure of the galaxy, such as the disk and spiral arms, produces the minimum possible alteration in the stream. These streams, see Figs.~\ref{fig:MWstreamsGC} and~\ref{fig:MWstreams}, have been extracted from \citet{10.1093/mnras/stad321}.  The intention in this section is to extract the value of the torsion of the parametrized stream curves with Eq.~(\ref{torsionfromders}) and check for their vanishing (or not).

We have taken two of the streams out of further consideration, namely those at Orphan-Chenab and Styx. The reason is that they may be influenced by gravity sources outside the MW. 
Due to the proximity of the Large Magellanic Cloud (LMC) to our galaxy, the streams in its periphery in the angular direction of that cloud could suffer alterations due to this additional source of gravity \citet{Conroy_2021}, \citet{Lilleengen_2022}.  
\textcolor{black}{Moreover, time-dependent perturbations such as those caused by the Magellanic cloud~\citet{2019MNRAS.487.2685E,2021MNRAS.501.2279V} difficult the interpretation of the stream track as that of an orbit, accruing further uncertainty (though they can be used in reverse to constrain the time-dependent potential of the perturbing body).
This means that even those streams not removed from the analysis could be affected by time-dependent perturbations that we do, at present, not capture.  Sometimes the effect can be very pronounced, such as a visible kink in the stream, others it can be more subtly  manifested by, for example, a misalignment between the stream track and the  velocity vectors~\cite{2021MNRAS.501.2279V}.
}

To obtain the torsion of the curves following the stream \textcolor{black}{we have  employed both a smooth polynomial parametrization and a (piecewise) elliptical-helicoidal parametrization that we here report. }
Therefore, we first fit each of the galactocentric Cartesian coordinates tracking the individual streams in the data compilation of~\citet{10.1093/mnras/stad321} to order-four polynomial or elliptical/helicoidal 
 parametric curves.  The parameter that describes each curve takes values in the interval $t\in[0,1]$. 
\textcolor{black}{The generic parametrizations are relegated to the appendix, see Eq.~(\ref{eq:MWreconstruction}) and following.} 
The parameters $k$ or $t$ therein are  arbitrary coordinates that can be converted to arc length, that has clearer geometric significance,
by means of 
\begin{equation}
	s = \int \sqrt{\sum^3_{i=1} \left(\frac{dx^i}{dt}\right)^2} dt\ ;
\end{equation}
the derivatives in Eq.~(\ref{torsionfromders}) are to be taken respect to the parameter $t$, in general, or $s$, if a change is variables is effected (the outcome is the same, of course) to obtain the torsion. 

To work with the streams in the database\footnote{\href{https://github.com/cmateu/galstreams}{https://github.com/cmateu/galstreams}} we use the \textsc{galstream} library \textcolor{black}{and to perform the fits we use a least-squares method with a standard python installation.} 

\textcolor{black}{
A word about the uncertainty in this extraction is warranted. 
The data points $(x, y, z)$ for the extracted stream trajectories are quoted without $(\Delta x, \Delta y, \Delta z)$ uncertainties  in the original reference~\citet{10.1093/mnras/stad321}. 
Possible large systematic errors, particularly due to distance tracks which are difficult to pinpoint with accuracy, can at present not be estimated.
Thus, the understood uncertainty of our parametric reconstruction stems in its entirety from our own fit interpreting the stream tracks in terms of Eq.~(\ref{ellipticalhelix}).  Until uncertainties in the track data on which we lean are compiled, the systematics will remain unclear. 
The reconstructed torsion will only become accurate when higher-precision estimates of the distance gradient along a stream with estimated uncertainties are provided.
}

After the reconstruction of the parametric curves we can obtain the expression for the torsion of each curve ${\bf r} = (x, y, z)$ using once more  Eq.~(\ref{torsionfromders})  $\displaystyle \tau = \frac{{\rm Det}({\bf r}', {\bf r}'', {\bf r}''')}{||{\bf r}' \times {\bf r}''||^2}$,
where the derivative is taken respect to the $t$ parameter, $'=d/dt$. We analytically express the derivatives in terms of the polynomial/elliptical parametrizations and then evaluate them as function of the $t$ parameter that we used for the fit,  taking values from 0 to 1. Because the torsion is parametrization independent, the torsion can also be given as function of the arclength
$\tau(s)$ calculating the derivatives respect to either of $t$ or $s$.

\begin{figure*}
  \centering
  \includegraphics[width=0.275\linewidth]{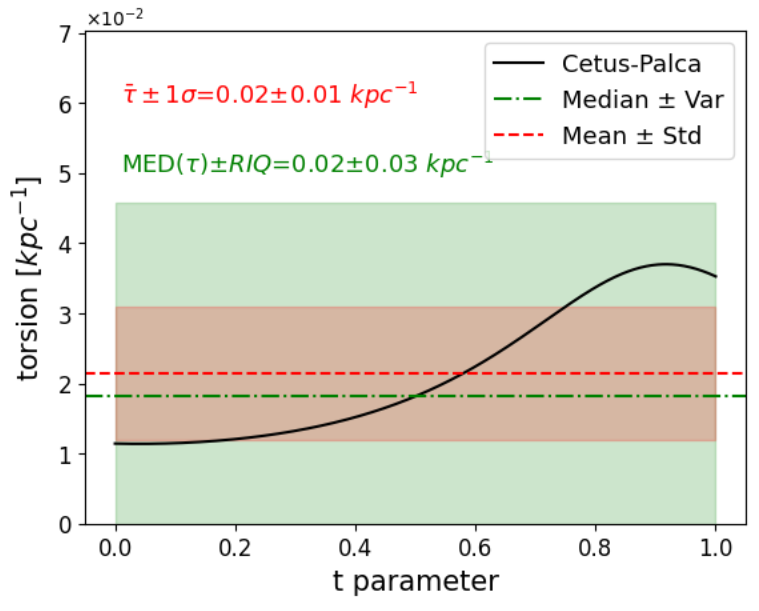}
  \includegraphics[width=0.275\linewidth]{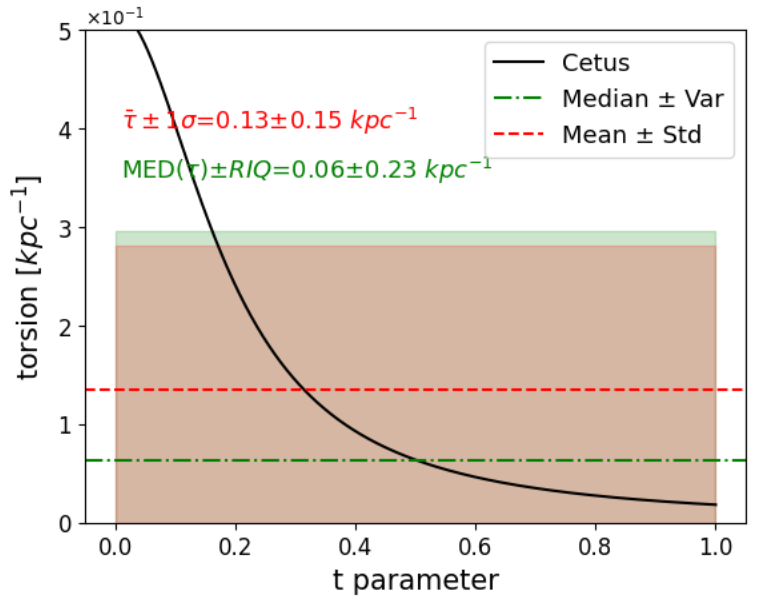} 
  \includegraphics[width=0.3\linewidth]{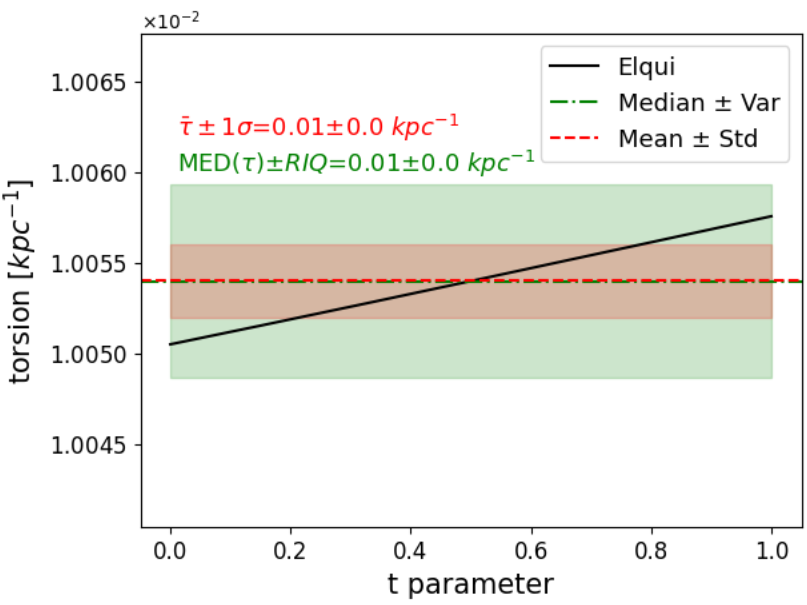}

  \includegraphics[width=0.3\linewidth]{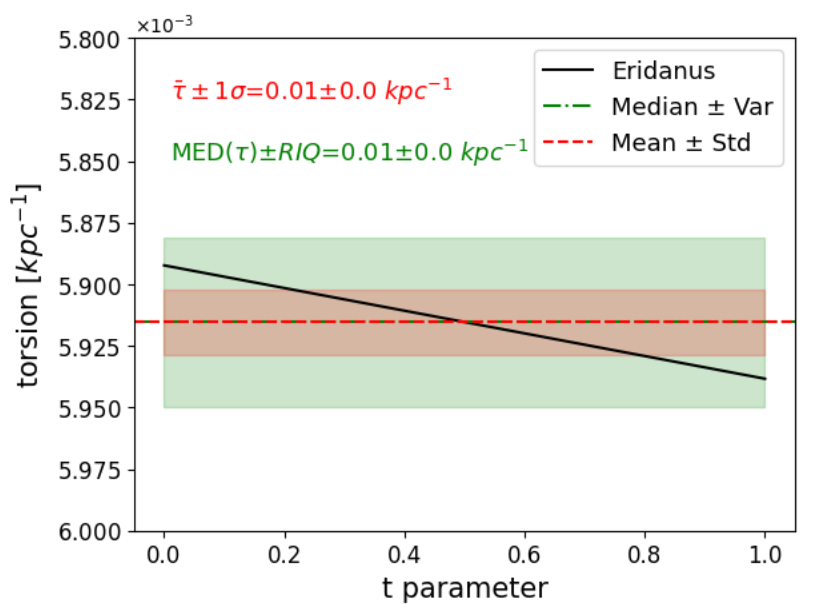}
  \includegraphics[width=0.3\linewidth]{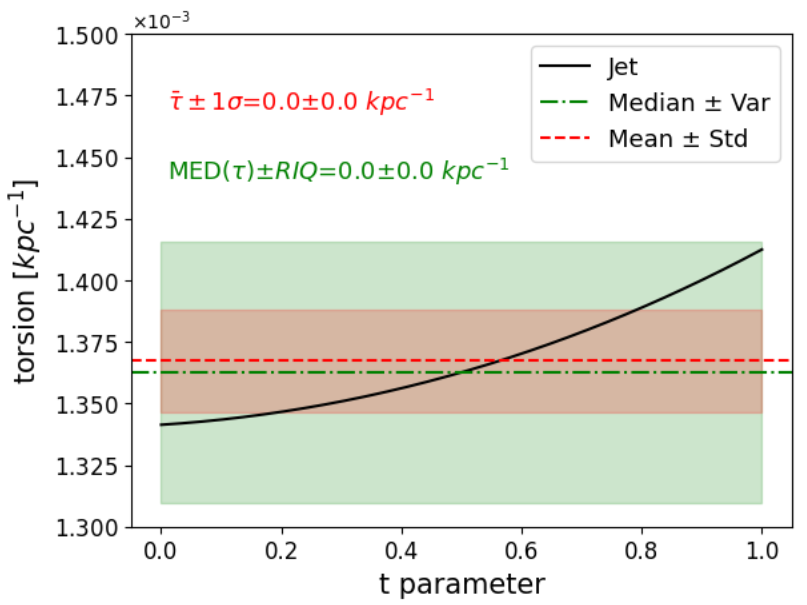}
  \includegraphics[width=0.3\linewidth]{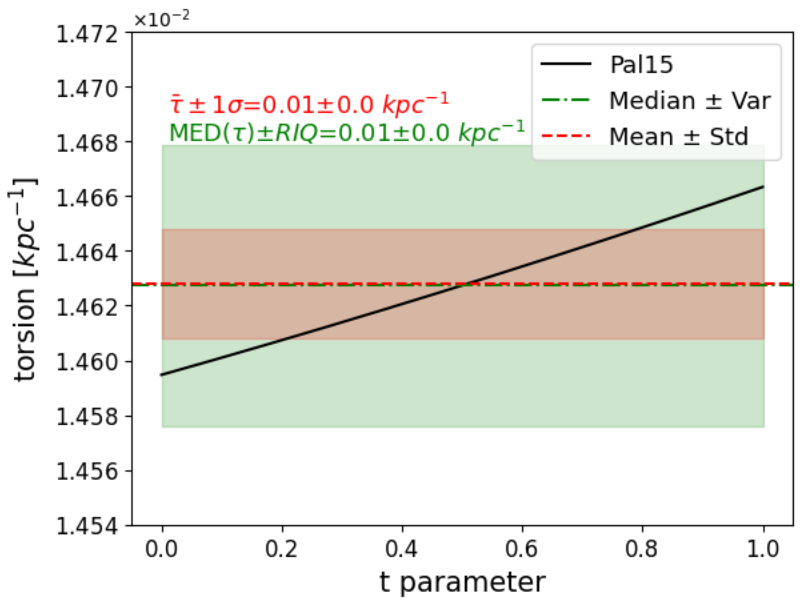}

  \includegraphics[width=0.3\linewidth]{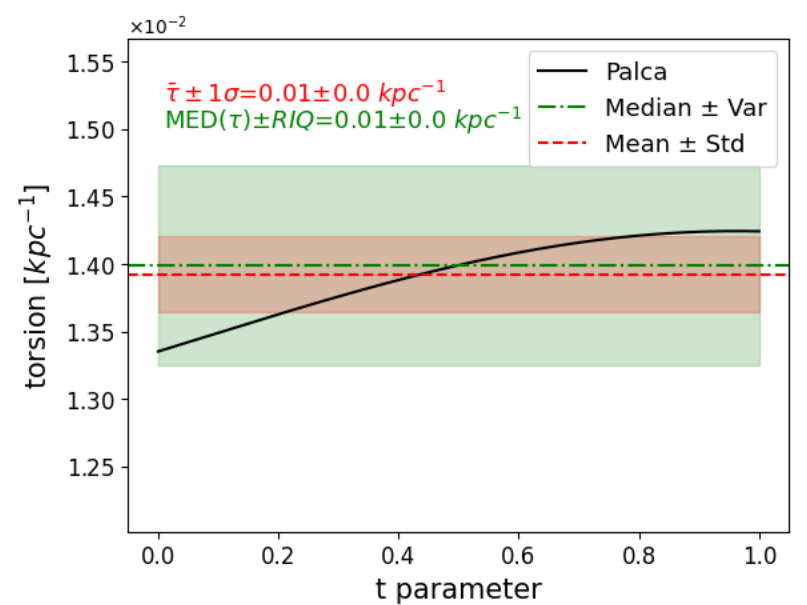}
  \includegraphics[width=0.275\linewidth]{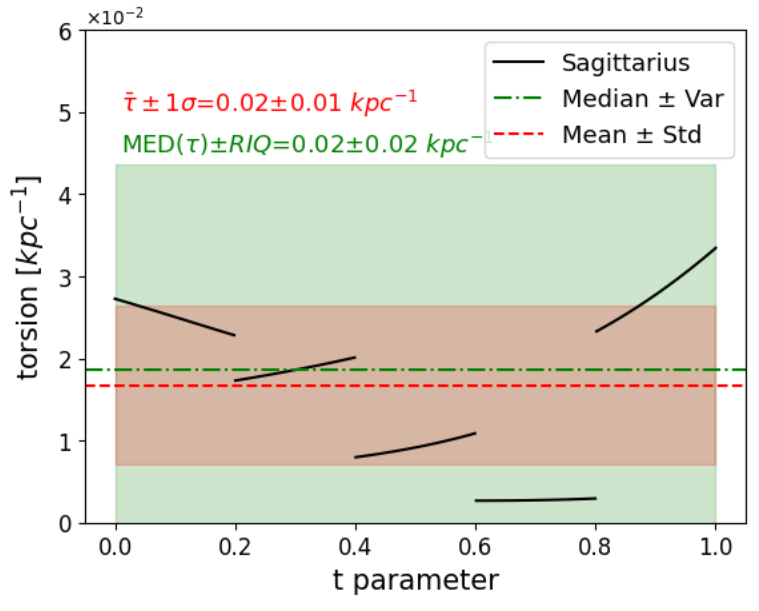}
  \includegraphics[width=0.3\linewidth]{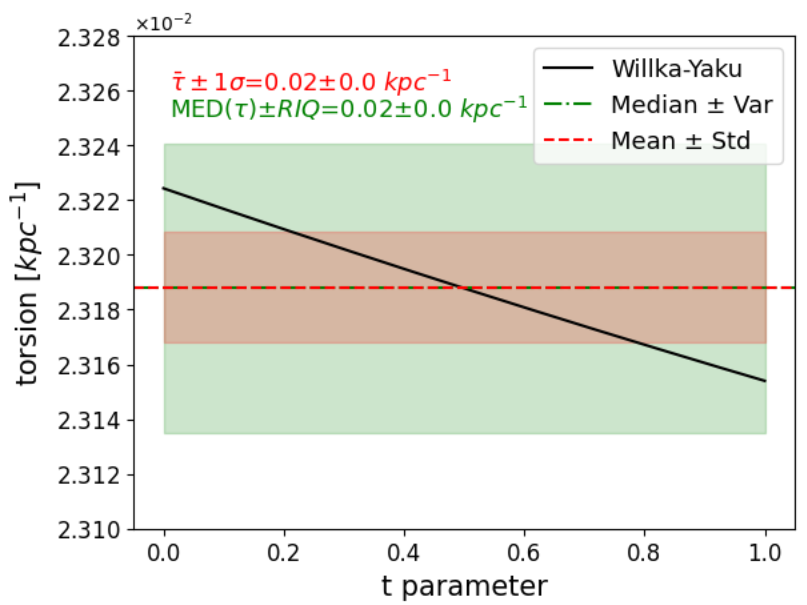}

  \caption{ \color{black} Local torsion $|\tau|$ of the MW streams considered in this work, calculated using Eq.~(\ref{torsionfromders}). The green and red lines and bands of the same shade represent the median value with the interquartile range and the mean value plus/minus one standard deviation, respectively. Stellar streams such as Eridanus and Jet show values of the torsion of order $\tau  \sim O(10^{-3})$, so their torsion is to be taken compatible with the influence of a source of the intensity of the galactic plane. The fit to  the Sagittarius data has been performed dividing the stream into 5 slices due to its irregular shape. }
  \label{fig:MWtorsion}
\end{figure*}

\textcolor{black}{The local torsion along many of the streams in Fig.~\ref{fig:MWtorsion} takes significant values, above the expected $O(10^{-4})$ ``noise'' that we found in the simulations when the streams evolved around spherical sources, as well as the $O(10^{-3})$ or less from simulations that included a galactic plane. Higher values for the torsion have been found for streams such as Cetus, Willka-Yaku, Cetus-Palca and Sagittarius} 

\textcolor{black}{In fact, the two streams with largest central value of the torsion (Willka-Yaku and Cetus) 
do not appear, by eyeballing the actual stream, to have such important helicity. In the case of Cetus, the large fit uncertainty makes it irrelevant, and the case of Willka-Yaku, whose torsion is still near the distribution of the other streams, deserves being revisited in the future because its shape is very different from our fit functions. 
}

\textcolor{black}{In Table~(\ref{tab:CV}) we quantify the torsion in the stream sample:  we quote the mean and median values of the torsion among the parametrized pieces of the stream, and also the standard deviation and the interquartile range of the torsion distribution.
The high values  $\tau \sim O(10^{-2})$ found for the torsion in several cases are compatible with  a non-spherical gravitational source, and also perhaps with the interaction with other gravitational sources different from the overall galactic field, so the observable appears to merit future study.}
\begin{table}
\caption{ \textcolor{black}{From left to right: values for the mean, standard deviation, median and interquartile range of the absolute value of the torsion in kpc$^{-1}$, and value of the $\chi^2$ per number of degrees of freedom of the fit. Recall that, from section~\ref{subsec:suelo} we expect that values above $10^{-3}$ kpc$^{-1}$ may signal a nonspherical DM distribution as the effect of the aspherical galactic plane is not strong enough. } }
	\label{tab:CV}
	\centering
	\begin{tabular}{ |c|c|c|c|} \hline
					    &                             &                     &                  \\ 
 		\textbf{Stream} & $(\bar{\tau} \pm   \sigma)\cdot 10^{-2} $  & $(med(\tau) \pm $IQR)$\cdot 10^{-2}$ & $\chi^2/N_{gdl}$ \\  
					    &                             &                     &                  \\ \hline
 		Cetus-Palca     & 1.82 $\pm$ 0.95 & 1.82 $\pm$ 2.76  &  1.16    \\  \hline
		Cetus           & 6.37 $\pm$ 14.65 & 6.38 $\pm$ 23.18 & 2.13 \\  \hline
		Elqui           & 1.01 $\pm$  0.00 & 1.01 $\pm$  0.00 & 0.01 \\  \hline
		Eridanus        & 0.59 $\pm$ 0.00 & 0.59 $\pm$ 0.00 &  0.02    \\  \hline
		Jet             & 0.14 $\pm$  0.00 & 0.14 $\pm$  0.00 & 0.28 \\  \hline
		Pal15           & 1.46 $\pm$  0.00 & 1.46 $\pm$  0.00 & 0.85 \\  \hline
		Palca           & 1.40 $\pm$ 0.03 & 1.40 $\pm$ 0.07 &  0.27    \\  \hline
		Sagittarius     & 1.85 $\pm$ 0.96 & 1.85 $\pm$ 2.50 &  -    \\  \hline
 		Willka-Yaku     & 2.32 $\pm$  0.00 & 2.32 $\pm$  0.00 & 0.06 \\ \hline
	\end{tabular}
	\end{table}

Because torsion (as curvature) has dimensions of inverse length, we would expect 
 stellar streams to perhaps show an inverse relation with respect to their distance from the galactic center, 
as defined in  Eq.~(\ref{torsionfromders}). 
Irrespectively, in a galaxy such as the Milky Way the torsion of galactic streams should have a characteristic scale of 
 $(10$ kPc$)^{-1}$. 
As per the discussion around Eq.~(\ref{tcylinder}), where we established 
$\tau\in\left[ -\frac{1}{\rho},\frac{1}{\rho}\right]$, our selection of streams at 30 kpc or more means that we 
would consider values of the torsion of order  0.01 in units of inverse kiloparsec to be sizeable and very different from zero, and table~\ref{tab:CV} shows several such. Also, as per the discussion below Eq.~(\ref{tplanesphere}), those with $\tau$ above 0.001 kpc$^{-1}$ could perhaps carry interesting information about the 
DM distribution.

\textcolor{black}{
Turning to the data, the torsions that we seem to observe in the MW streams show mean values above $0.001$ that could give an insight into the DM distribution, but also information about the interaction of the MW with other gravitational sources. Several streams show a significant value for the torsion: only Eridanus and Jet seem to have torsion compatible with being seeded by the galactic plane, as small as $\tau \sim O(10^{-3})$. }

\textcolor{black}{
From the information we obtain from the streams in the database we cannot explore specific details on the shape of the gravitational source, beyond establishing its probable non-sphericity. 
}


\section{Conclusions and outlook}\label{sec:conclusion}

The problem of galactic rotation curves suggests that galaxies are surrounded by significant amounts of dark matter, and the overall shape of these sources is yet to be ascertained. Whereas spherical DM distributions around galaxies have to be fine-tuned to explain the flatness of rotation curves,  a cylindrical (or, generally, prolate) DM source can naturally explain that flattening. This avoids the fine-tuning of spherical DM haloes to precisely follow the $1/r^2$ fall-off for a large swath of $r$ values.
Observables inside the galactic  plane cannot however distinguish between spherical (though fine tuned) and cylindrical/elongated gravitational sources, but out-of-galactic-plane information could provide new strong discriminants.

The stellar streams around the Milky Way have been extensively investigated for a while now, and are still nowadays a relevant subject of research. \textcolor{black}{The track left by a stream, once extracted, can be used as the best proxy we have for a long-range trajectory of size commensurate to a galaxy, with all the attending dynamical uncertainties that this identification causes~\citet{2014ApJ...795...95B}, and thus a tool to infer the geometry of the impacting gravitational sources. }
Stream tracks can be characterized by their torsion according to Eq. ~(\ref{torsionfromders}); around a central potential orbits move in a plane and are expected to be torsionless (see Fig.~\ref{fig:cloudspherical}). In addition, test masses around cylindrical sources are expected to follow helical orbits in which the torsion is non-zero (see Fig.~\ref{fig:cylindrical}) if given a vertical velocity. Another approach is to consider an ellipsoid-shaped halo, which is not perfectly cylindrical but rather elongated. The expected orbit of the stream components would arise from the combination of the orbits around central potentials and the helical orbits around cylinders, as is seen in Fig.~\ref{fig:sphere+cloudcylinder}.
\textcolor{black}{Torsion is then an observable directly tailored to assess
the prolateness of a DM halo, as opposed to 
a multiparameter reconstruction which, though sensitive to the halo shape~\citet{2023MNRAS.521.4936K}, is affected by a degeneracy with many other features}.

The streams of the Milky Way have been a subject of research for a considerable time span, and many of the objects that constitute these streams have been catalogued. From a reconstruction of the stream tracks we infer the torsion caused by the gravitational source in Fig.~\ref{fig:MWtorsion}. In this work we only consider those streams that seem to be far enough away from the galactic center to (1) avoid large effects from the baryonic component of the galaxy and (2) have a bird's eye view of the DM halo from outside a large fraction thereof.  From the extraction of the torsion we see that it is non-negligible in some of the streams considered.

From our evaluation of the torsion we do not dare favor one or another interpretation of the DM halo shape in view of current data; this article should be seen as a proposal for a new observable, $\tau$, and a first exploratory study.
\textcolor{black}{We do find streams with significant torsion, which is encouraging and suggests that further scrutiny could eventually inform us about the shape of the halo.
We have payed no attention to the sign of $\tau$ since it only distinguishes left- from right-handed tracks. That opens another interesting study in itself, as, with the predominant galactic rotation determining a sign, the preference of a given handedness would inform an asymmetry in the vertical motion. But this would require much more statistics and take us far afield at present.
}

\textcolor{black}{The most trustworthy streams should be those stemming from kinematically cold globular cluster (low velocity dispersion), as those present the best defined tracks.}
In future observational work it might be interesting to actively seek streams that show both vertical motion (along the axis perpendicular to the MW plane) and also azimuthal motion around that axis, as those with large $v_z$ and $v_\varphi$ will
be most sensitive to the torsion. Should those streams show trajectories that are compatible with lying on a plane (zero torsion), a spherical halo would be preferred. Should they however appear helicoidal, with nonnegligible torsion, they would be pointing to an elongated DM halo.

Streams that may be detected in nearby galaxies carry the same information about their respective haloes.

\textcolor{black}{
However, for galactic-halo population studies with streams, which will require a 10-MPc reach, three-dimensional reconstruction of the streams appears more difficult; in that case, only the projected curvature is measurable \citet{2023ApJ...954..195N}, but not the torsion, and what can be learnt about the DM halo hangs on statistical analysis to reduce the impact of degeneracies due to the projection.
}

Finally, other observables can bear on the overall shape of the halo, and we are investigating 
the shape-sensitivity of the gravitational lensing of both electromagnetic and gravitational radiation.

\newpage
\begin{acknowledgements}
We thank our engineer David Fernández Sanz for maintaining an adequate computing environment suited to our needs at the theoretical physics departmental cluster, and Pablo Blanco-Mas for insightful conversations.
Financially supported by spanish ``Ministerio de Ciencia e Innovaci\'{o}n: Programa Estatal para Impulsar la Investigaci\'{o}n Cient\'{i}fico-T\'{e}cnica y su Transferencia'', Grants PID2021-124591NB-B-C41 and PID2019-108655GB-I00 funded by MCIN/AEI/ 10.13039/501100011033 and by the EU-ERDF, as well as Univ. Complutense de Madrid under research group 910309 and the IPARCOS institute. 
\end{acknowledgements}


\bibliography{ABQFLE.bib}

\newpage

\section{Appendix}
\label{section:apendix}
\subsection{Track fits}

\color{black}A simple 
 parametrization $(x(t), y(t),z(t))$ that we have employed is a piecewise fourth-order polynomial fit reading 
\begin{equation}
	x^i (t) = a_i t^4 + b_i t^3 + c_i t^2 + d_i t + e_i \thinspace \thinspace \thinspace \text{for i=1,2,3}.
	\label{eq:MWreconstruction}
\end{equation}
It is idle to try to relate $t$ to a Newtonian time since, not knowing {\it a priori} the dynamics of the system, it is unknown at which time a star was at what position along its trajectory. Only the instantaneous (present) geometry of the stream is known with certainty, and therefore an arbitrary parameter $t$ (or the arc length after computing it) should suffice.

Another simple yet useful parametrization is that of an elliptical helix, 
\begin{eqnarray} \label{ellipticalhelix}
\nonumber    {\bf x}(t) &=& {\bf x}_0 +
    \left(
    a\ \cos(\omega\ t+\phi), b\ \sin(\omega\ t+\phi),ct
    \right)\\ \nonumber
    {\bf x}'(t) &=& 
    \left(
    -a\omega\ \sin(\omega\ t+\phi), b\omega\ \cos(\omega \ t +\phi),c \right)\\ \nonumber
    {\bf x}^{''}(t) &=& 
    \left(
    -a\omega^2\ \cos(\omega\ t+\phi), -b\omega^2\ \sin(\omega\ t+\phi),0 \right)\\ \nonumber
        {\bf x}^{'''}(t) &=& 
    \left(
    a\omega^3\ \sin(\omega\ t+\phi), -b\omega^3\ \cos(\omega\ t +\phi),0 \right) \\
\end{eqnarray}
with $(a,b,c,\omega, \phi,\textbf{x}_0)$ the parameter set such that for $t=0$ the curve cuts the $XY$ plane
with $z=0$ with initial phase $\phi$ in the $x-y$ rotation, $c$ controlling the advance of the helix
and $a$, $b$ the elliptic projection on the $XY$ plane. Should the rotation axis also need to be chosen 
different from $OZ$ an additional orthogonal matrix $O(\varphi,\theta,\xi)$ parametrized by three Euler angles
should be introduced as ${\bf x}\to O{\bf x}$. This we do not employ here out of simplicity, assuming the $OZ$
axis is an axis of symmetry for the DM halo and coincides with the galactic rotation axis.

We need to distinguish two types of fits. 

First, when fitting our simulated data, there is no obstacle in choosing the coordinate axes so that  $OX$ lies along the initial visual from the galactic center to the globular cluster which will give rise to the stream.
In that case, the offset ${\bf x}_0$ in Eq.~(\ref{ellipticalhelix}) is unnecessary.

For the purposes of the fit, the elliptical parametrization is then, eliminating time,
\begin{eqnarray}
x(z) &=& A\cos\left(
\frac{\omega}{C} (z-z_0) + \phi \right) + x_0\\
y(z) &=& B \sin\left(
\frac{\omega}{C} (z-z_0) + \phi \right) + y_0
\end{eqnarray}
the length parameters \textcolor{black}{$A$ and $B$ take values proportional to the distance of the stream to the galactic center in kpc. $C$ is a vertical ``velocity'' and $\omega$ an ``angular velocity'' in the $XY$ plane; but because of the arbitrariness of the parameter $t$, they are only relevant in the combination $\omega/C$ that indicates the angle advanced per unit height in the corkscrew motion characterizing torsion. }

(If one insists in thinking of the orbit of a single star instead of the track of a stream, then
the angular velocity is of the order of magnitude
$
\omega \sim \frac{v_\perp}{r_\perp} \sim 
\frac{220 {\rm km/s}}{10^4 \rm Pc} \times 3.15 \ 10^{16}\ {\rm s}/{\rm Gyr}\times 
\frac{1}{3.1\ 10^{13} {\rm km}/{\rm Pc}}\ ,
$
resulting in $\omega\sim 22$ rad/Gyr.)

The second type of fit involves real data. Then, the osculating plane for the starting point of each piece of the orbit need not be the aligned with the galactic Cartesian axes in account for initial conditions and precession in a noncentral field. We thus need a minimum of one additional angle $\Theta$ for each fit, to orient that starting plane,
\begin{eqnarray}
    x' &=& x \cdot \cos{\Theta} - y\cdot \sin{\Theta}  \\
    y' &=& x \cdot \sin{\Theta} + y\cdot \cos{\Theta} . 
\end{eqnarray}
Additionally, real streams may need an offset ${\bf x}_0=(x_0,y_0,z_0)$ bringing the total number of fit parameters to nine.

Once the parametrization has been chosen, we need to decide what is the optimal track through a cloud of them in a simulation. A strategy is to employ a squared-distance minimization strategy.
The function to be minimized is the following sum running over each of the stars in the cloud,
\begin{equation}
    D^2(a,b,c,t_0,\omega,t_i) = \sum_{i=1}^N ({\bf x}_i-{\bf x}(t_i))^2
\end{equation}
where $t_i$ will be the point along the curve for which its distance to the $i$th point will be minimized. Taking the partial derivatives the set of equations
\begin{eqnarray}
  \sum_{i=1}^N ({\bf x}_i-{\bf x}(t_i))\cdot {\bf v}(t_i) =0 \\
  \sum_{i=1}^N ({\bf x}_i-{\bf x}(t_i))\cdot \frac{\partial {\bf x}(t_i)}{\partial a_j}
\end{eqnarray}
constrain the wanted parameters $a_j$ chosen among the fittable set. The first of these equations is the condition of orthogonality of the curve's tangent to the visual to the point.

Small stretches of a stellar stream can be fit with the functional form of Eq.~(\ref{ellipticalhelix}) for a straightforward extraction of the torsion. 
The torsion for this elliptical helix is then 
\begin{equation}
\tau= \frac{abc\omega}{b^2c^2\sin^2(\omega\ t + \phi) + a^2c^2\cos^2(\omega\ t +\phi)+a^2b^2\omega^2}.
\end{equation}
This is much simplified and becomes a $t$-independent constant for a circular helix with $a=b$, as given in the main text in Eq.~(\ref{torsionestimate}).  

\subsection{Equations of motion for $N$-body simulations}
Finally, we specify the (simplest available)  numerical method employed to simulate an $N$-body stellar stream in section~\ref{subsec:simulation1}.
 The positions of the stellar objects composing it are updated in Cartesian coordinates. 
For this we  use Euler's Method with time step $\Delta t=t_f/N_t$, with the velocity updated via a once-improved Euler step, 
\begin{eqnarray}
	x_{j+1}^i&=&x^i_j + \Delta t v^i_j \thinspace \thinspace \thinspace \thinspace \text{for i=1,2,3} \\
       v^i_{j+1}&=&v^i_j + \frac{1}{2}\Delta t \space f^i({\bf x}_j+\frac{1}{2}h{\bf v}_j) \thinspace \thinspace \thinspace \thinspace \text {for i=1,2,3}
\end{eqnarray}
where $f^i$ is the function yielding each component's acceleration.

The acceleration is calculated at each step from standard formulae such as 
\begin{eqnarray}
a_{i=1,2,3}= &-&GM\frac{x^i}{(x^2+y^2+z^2)^{3/2}} \nonumber\\
             &-& \sum^{N-1}_{j=1}G\frac{m_jx^i}{((x-x_j)^2+(y-y_j)^2+(z-z_j)^2)^{3/2}}
	\label{eq:accelerationspherical} 
\end{eqnarray}
for a spherical central source.
The first line of this expression is the acceleration caused by the central spherical source, and the second is the force that attempts to bind the stellar-stream stars together (and that is too weak to avoid the tidal stretching). 

If we turn to a cylindrical source, the force external to the stream in Eq.~(\ref{eq:accelerationspherical}) needs to be replaced, so that 
\begin{eqnarray}
	a_{i=1,2}=   &-& 2G\lambda\frac{x^i}{x^2+y^2}  \label{eq:cylacc} \\
                                &-& \sum^{N-1}_{j=1}G\frac{m_jx^i}{((x-x_j)^2+(y-y_j)^2+(z-z_j)^2)^{3/2}} \nonumber\\
	a_{i=3}    =   &-& \sum^{N-1}_{j=1}G\frac{m_jx^i}{((x-x_j)^2+(y-y_j)^2+(z-z_j)^2)^{3/2}} \label{eq:cylacc2} .
\end{eqnarray}
At last, we have simulated both a cylinder+sphere  and a plane+sphere background fields, with respective forces given in Eq.~(\ref{cylsph}) and~(\ref{accsphereplane}). 
%
%

\end{document}